\documentclass{aa}  
\usepackage{color}
\usepackage[varg]{txfonts}
\usepackage{sidecap}
\usepackage[stable]{footmisc}
\usepackage{lscape}
\usepackage{graphicx}
\usepackage{natbib,twoopt}
\usepackage{subcaption}
\bibpunct{(}{)}{;}{a}{}{,}
\usepackage[breaklinks=true]{hyperref} %% to avoid \citeads line fills
\bibpunct{(}{)}{;}{a}{}{,}             %% natbib format for A&A and ApJ
\makeatletter
  \newcommandtwoopt{\citeads}[3][][]{\href{http://adsabs.harvard.edu/abs/#3}%
    {\def\hyper@linkstart##1##2{}%
     \let\hyper@linkend\@empty\citealp[#1][#2]{#3}}}
  \newcommandtwoopt{\citepads}[3][][]{\href{http://adsabs.harvard.edu/abs/#3}%
    {\def\hyper@linkstart##1##2{}%
     \let\hyper@linkend\@empty\citep[#1][#2]{#3}}}
  \newcommandtwoopt{\citetads}[3][][]{\href{http://adsabs.harvard.edu/abs/#3}%
    {\def\hyper@linkstart##1##2{}%
     \let\hyper@linkend\@empty\citet[#1][#2]{#3}}}
  \newcommandtwoopt{\citeyearads}[3][][]%
    {\href{http://adsabs.harvard.edu/abs/#3}
    {\def\hyper@linkstart##1##2{}%
     \let\hyper@linkend\@empty\citeyear[#1][#2]{#3}}}
\makeatother 
\begin{document}

   \title{Single stars in the Hyades open cluster\thanks{Based on observations collected at the Centro Astron\'{o}mico Hispano Alem\'{a}n (CAHA) at Calar Alto, operated jointly by the Max-Planck Institut f\"{u}r Astronomie and the Instituto de Astrof\'{i}sica de Andaluc\'{i}a (CSIC)}}

   \subtitle{Fiducial sequence for testing stellar and atmospheric models}

\author{Taisiya G. Kopytova\inst{1,2} 
\and Wolfgang Brandner\inst{1}
\and Emanuele Tognelli\inst{3,4}
\and Pier Giorgio Prada Moroni\inst{4,5}
\and Nicola Da Rio\inst{6}
\and Siegfried R\"{o}ser\inst{7}
\and Elena Schilbach\inst{7,1}
}

\institute{Max-Planck-Institut f\"ur Astronomie, K\"onigstuhl 17, 69117 Heidelberg, Germany\\
e-mail: kopytova@mpia.de
\and International Max-Planck Research School for Astronomy and Cosmic Physics at the University of Heidelberg, IMPRS-HD, Germany
\and University of Roma Tor Vergata, Department of Physics, Via della Ricerca Scientifica 1, 00133, Roma, Italy
\and INFN, Section of Pisa, Largo Bruno Pontecorvo 3, 56127, Pisa, Italy
\and University of Pisa, Department of Physics 'E.Fermi', Largo Bruno Pontecorvo 3, 56127, Pisa, Italy
\and Department of Astronomy, University of Florida, 211 Bryant Space Science Center, Gainesville, FL 32611, USA
\and Astronomisches Rechen-Institut, Zentrum f\"{u}r Astronomie der Universit\"{a}t Heidelberg,
M\"{o}nchhofstr. 12-14, 69120 Heidelberg, Germany}

\date{Received / Accepted}

\abstract{Age and mass determinations for isolated stellar objects remain model-dependent. While stellar interior and atmospheric theoretical models are rapidly evolving, we need a powerful tool to test them. Open clusters are good candidates for this role.}
{We aim to create a fiducial sequence of stellar objects for testing stellar and atmospheric models.}{We complement previous studies on the Hyades multiplicity by Lucky Imaging observations with the AstraLux Norte camera. This allows us to exclude possible binary and multiple systems with companions outside 2--7 AU separation and to create a "single-star sequence" for the Hyades. The sequence encompasses 250 main-sequence stars ranging from A5V to M6V. Using the "Tool for Astrophysical Data Analysis" (TA-DA), we create various theoretical isochrones applying different combinations of interior and atmospheric models. We compare the isochrones with the observed Hyades single-star sequence on \mbox{$J$ vs. $ J - K_s$}, $J$ vs. $J - H$  and $K_s$ vs. $H - K_s$ color-magnitude diagrams. As a reference we also compute absolute fluxes and magnitudes for all stars from X-ray to mid-infrared based on photometric measurements available in the literature(ROSAT X-ray, GALEX UV, APASS $gri$, 2MASS $JHK_s$, and WISE $W1$ to $W4$).}{We find that combinations of both PISA and DARTMOUTH stellar interior models with BT-Settl 2010 atmospheric models describe the observed sequence well. We use PISA in combination with BT-Settl 2010 models to derive theoretical predictions for physical parameters (Teff, mass, log $g$) of 250 single stars in the Hyades. The full sequence covers the mass range 0.13 to 2.3 Msun, and effective temperatures between 3060 K and 8200 K.}{Within the measurement uncertainties, the current generation of models agree well with the single-star sequence. The primary limitations are the uncertainties in the measurement of the distance to individual Hyades members, and uncertainties in the photometry. Gaia parallaxes, photometry and spectroscopy will greatly reduce the uncertainties in particular at the lowest mass range, and will enable us to test model predictions with greater confidence. Additionally, a small ($\sim$0.05 mag) systematic offset can be noted on $J$ vs. $J - K$ and $K$ vs. $H - K$ diagrams - the observed sequence is shifted to redder colors from the theoretical predictions.}


\maketitle
\authorrunning{Kopytova et al.}

%
%________________________________________________________________

\section{Introduction}

The knowledge of physical parameters of stellar and sub-stellar objects provides the basics for astrophysics.
Observations allow us to determine distance (using parallaxes), age (through membership in clusters or moving groups), mass (by calculating orbital parameters in binary systems), and radius (in transiting and eclipsing systems). Asteroseismic scaling relations can also provide estimates for stellar mass and radius. However, these relations mays suffer from unknown systematics effect and require additional calibration using observations and theoretical models \citep[e.g.][]{Miglio_etal_2012}. Therefore, when dealing with isolated stellar objects, we usually have to rely on theoretical models to determine mass and age.
While stellar and atmospheric models are rapidly evolving, we need a powerful tool to test and calibrate them.
Open clusters are good candidates for this role, since they contain many coeval objects of the same chemical composition spanning a range of masses,
thus avoiding the problems of small number statistics.
Additionally, open clusters have distance and age estimates that are independent of theoretical models to be tested.

In general, the validation of theoretical models consists of two independent problems - one is to test stellar interior structures, another is to inspect atmospheric models.
However, this process is non-trivial because interior structure models provide us with physical parameters, such as effective temperature, luminosity, mass and age, that need to be converted into observed quantities (magnitudes, colors and fluxes). This conversion exploits grids of synthetic spectra provided by atmospheric models and requires careful calibration for specific photometric bands \citep[e.g.][]{Da_Rio_Roberrto_2012}.

In the past, open clusters have already been used as benchmarks for theoretical models. \citet{Bell_etal_2012} used the Pleiades to test existing pre-main sequence isochrones
by comparing their predictions with well-calibrated color-magnitude diagrams in the wavelength range of 0.4--2.5 $\mu$m.
Bell et al. have shown that no pre-main sequence model can describe the observed Pleiades sequence for the temperatures cooler than 4000 K.
The predicted fluxes are over-estimated by a factor of 2 at 0.5  $\mu$m, with the difference decreasing with increasing wavelength.

In comparison to the Pleiades \citep[d $\sim$ 120--140 pc;][]{Percival_etal_2005, van_Leeuwen_2009}, the Hyades open cluster is closer to the Sun \citep[d $\sim$ 45 pc;][]{Perryman_etal_1998, van_Leeuwen_2009}.
This allows us to resolve companions with smaller physical separations, and also to analyze less luminous, hence less massive, objects.
Moreover, \citet{Roeser_etal_2011} reported 724 likely members of the Hyades, with individual kinematic distance estimates using the convergent point method \citep[e.g.][]{van_Leeuwen_2009}.
Individual distance measurements allow us to get more precise absolute magnitudes for each member of the Hyades.
The previous attempt to test stellar models with the Hyades \citep{Castellani_etal_2001} has shown a discrepancy between theoretical predictions and the observed main-sequence sample, especially at the region of the coolest stars. 

In this paper, we present a fiducial "single-star sequence" in the Hyades based on literature data and our own AstraLux Lucky Imaging observations.
Further, we use the obtained sequence to provide a test for commonly-used stellar and atmospheric models.
The paper is organized as follows: Section 2 describes observations and data from the literature; Section 3 discusses the theoretical models used;
in Section 4 we show the "single-star sequence" in the Hyades compared to theoretical isochrones calculated using various stellar, and atmospheric models and determine physical parameters of each Hyades "single-star"; in Section 5 we discuss our results and give a brief summary.

%__________________________________________________________________

\section{Observations and literature data}
To test interior and atmospheric models we use $JHK_{s}$ 2MASS data \citep{Skrutskie_etal_2006} which allows us to get a homogenous photometric set for all the 724 possible members of the Hyades. The angular resolution of 2MASS is limited to $\sim 3\arcsec$. To avoid a scatter on color-magnitude diagrams that can be introduced by unresolved binary and multiple systems, or optical blends with unrelated field or background stars, we check the literature and archived data bases for multiplicity and perform Lucky Imaging observations. The saturation threshold of the 2MASS photometry is $K_{s} \sim 4$ mag, therefore fluxes for the brightest stars are replaced by the data from \citet{Carney_1982}, with a transformation to the 2MASS system applied. 
The color uncertainties are calculated in the standard way exploiting independent photometric magnitude uncertainties estimates.

\subsection{Literature data}
We check previous studies that make an attempt to identify binary and multiple systems in the Hyades including works of  \citet{Patience_etal_1998}, \citet{Mermilliod_etal_2009}, \cite{Morzinski_2011}, \citet{Duchene_etal_2013} and archival data from Hipparcos, Hubble Space Telescope (HST) and Washington Double Star \citep[WDS;][]{Mason_etal_2001} catalogs. The HST data have also been previously studied by \cite{Gizis_Reid_1995}, \cite{Reid_Gizis_1997} and \cite{Reid_Mahoney_2000}. Characteristics of the surveys are summarized in Table \ref{mult_table}.

\begin{table*}
\begin{center}

	\begin{tabular}{r||ccccc}
	Survey & Telescope & Method & Angular Resolution & Observed &  Binary/Multiple \\
	\hline
	 &&&&&\\
	Patience et al. (1998) & 5m Hale telescope & 2.2 $\mu$m  &0$\farcs$11 & 163 &33\\
				           & Palomar observatory & speckle imaging&&&   \\
				           &&&&&\\
	Mermilliod et al. (2009)  & Swiss 1m telescope & CORAVEL & - & 139& 25\\
				           & Haute-Provence Observatory & spectroscopy &&&\\
				           &&&&&\\
				           Morzinski (2011) & Keck and Lick & Adaptive & 0$\farcs$06&75&30\\
				           & observatories & optics &&& \\
				           &&&&&\\
				           Duch\^{e}ne et al. (2013) & Keck II & Adaptive & 0$\farcs$06&9&5\\
				           &&optics&&&\\
				           &&&&&\\

				           HST &Hubble Space&Imaging&0$\farcs$05&57&8\\
				            &Telescope&&&&\\
				             &&&&&\\
				           Hipparcos catalogue &Hipparcos Space&Imaging&0$\farcs$10& 195 &22\\
				            &Mission&&&&\\
				             &&&&&\\
				            AstraLux Norte& 2.2m telescope& Lucky &0$\farcs$11&198&40 \\
				            & Calar Alto observatory & imaging &&& \\
				               &&&&&\\
	\end{tabular}
	\end{center}
	\caption{Characteristics of various multiplicity surveys in the Hyades. Some stars were studied by several surveys. The WDS catalog is not presented in the table because its instrument characteristics is inhomogeneous.}
	
\label{mult_table}	
\end{table*}

\subsection{AstraLux lucky imaging observations}
The Lucky Imaging technique is based on series of short (few to several 10 ms) exposures, which "freeze" the speckle clouds caused by atmospheric turbulence. The series of frames is sorted by the best quality (based on the brightest pixel), and typically the best 1\% to 10\% of the images are shifted and co-added, resulting in a close to diffraction limited image.
AstraLux Norte is a lucky imaging camera mounted on the 2.2m telescope at the Calar Alto observatory in Spain \citep{Hormuth_etal_2008}.
The camera has a field of view of  $24\arcsec \times 24\arcsec$ (FOV) and a pixel scale of  $\sim$ 47 mas/px.

The observations were performed in Nov. 2011, Nov. and Dec. 2012. 
Depending on the brightness of the target and observing conditions, either 15 or 30 ms exposure time was chosen, so that 20,000 or 10,000 short exposures were obtained for each target, respectively. Only one quadrant of the detector was read-out to facilitate shorter integration times.

\begin{figure}
	\centering
	
		\includegraphics[width=8cm]{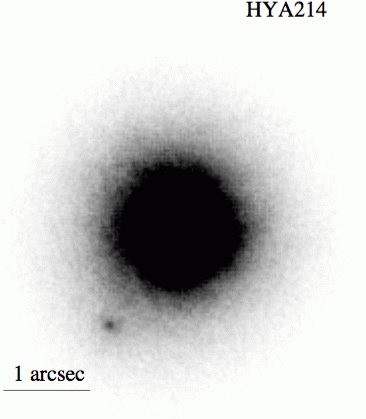}

	\caption{AstraLux Norte $z'$-band image of a binary candidate in the Hyades. The image is shown in log scale.}
\label{al_frame}
\end{figure}

The Hyades targets for the AstraLux run are pre-selected based on the results of the literature and archive check. In total 198 Hyades members are observed in SDSS $i'$ and $z'$ filters. The analysis for each target is done based on the best 1$\%$ frames that are co-added together. Possible companions are identified by a visual check of each co-added frame (see Fig. \ref{al_frame}) resulting in 40 Hyades members that reveal one or several companion candidates. 28 of the 40 systems with companion candidates did not have any binary/multiplicity record in previously published surveys.
Most of the AstraLux candidates to binary and multiple systems have only one epoch of observations, therefore they cannot be confirmed as common proper motion companions. Non-detections exclude companions outside a projected separation of 3--7 AU (depending on a distance to a star), though these stars could still have unresolved companions closer-in.

\vspace{12pt}
In total 463 Hyades members are observed in one or more surveys and 213 of them are identified as candidates for binary or multiple systems leading to 250 stars that do not have identified companions outside projected separation of 2--7 AU (depending on survey and distance to the star). Importantly, the angular resolution of instruments in the surveys we use varies from 0$\farcs$05 to 0$\farcs$11, which, additionally to various stars' distances, gives a range of separation where companions can be detected (2--7 AU; see \mbox{Table 2}).
Hence, this means different detection thresholds for companions  depending on in which survey(s) the system was observed.
Since this complicates the statistical analysis for constraining the number of binary and multiple systems, a robust estimate of the multiplicity properties of the Hyades will be a subject to a separate paper.

\section{Models}
In this section we describe the stellar evolutionary models and synthetic spectra used to produce theoretical isochrones for the Hyades and compare them with observed single-star sequence.

\subsection{Interior models}
We use two different stellar interior models - DARTMOUTH and new PISA grids produced for the Hyades case. 
Most recently, new BHAC15 \citep{Baraffe_etal_2015} stellar models have been published. 
These models are adherent to the widely-used BCAH98 models \citep{Baraffe_etal_1998}.
However, BHAC15 models do not provide interior structures for super-solar metallicities yet, therefore, we do not use them for the isochrone comparison.

The Dartmouth Stellar Evolution Program (DSEP)  \citep{Dotter_etal_2008} is a family of stellar evolutionary models exploring a wide range of metallicities and $\alpha$-enhancements. The basic equation of state for tracks $M \ge 0.8 M_{\odot}$ is a general ideal gas EOS with Debye-H\"{u}ckel correction \citep{Chaboyer_Kim_1995}. The FreeEOS in the EOS4 configuration is used for the lower mass tracks. Evolutionary tracks are computed for masses from 0.1 to 4 $M_{\odot}$. We use isochrones for [Fe/H] = 0.14 and [$\alpha$/H] = 0.0. For the Hyades metallicity, the DSEP models include a convective core overshooting of 0.2 times pressure scale height for \mbox{M $\ge 1.1$ M$_\odot$}.

The new PISA models have been computed by means of the FRANEC stellar evolutionary code \citep{Degl'Innocenti_etal_2008}, adopting input physics similar to those already discussed in \citet{Tognelli_etal_2011} and \citet{Dell'Omodarme_etal_2012}. 
The main difference with respect to the models already available on the Pisa database page\footnote{\url http://astro.df.unipi.it/stellar-models/} is the adoption of the SCVH95 EOS for the computation of stellar models with mass lower than 0.2 $M_{\odot}$. 
We adopt the recent \citet{Asplund_etal_2009} solar metal distribution, and the corresponding mixing length parameter calibrated on the Sun, namely $\alpha_\mathrm{ML} = 1.74$.  We also include a mild convective core overshooting ($\beta_\mathrm{ov} = 0.2$) for M $\ge 1.2$ M$_\odot$ \citep{Tognelli_etal_2012}. The models have been computed in the mass range [0.1, 2.8] M$_\odot$ from the early pre-MS evolution up to the exhaustion of the central hydrogen. The corresponding isochrones in the age interval [400, 800] Myr, with an age spacing of 10 Myr, have been generated. The stellar models have been computed for [Fe/H] = $+0.14$, which adopting the \citeauthor{Asplund_etal_2009} $(Z/X)_\odot = 0.0181$ and $\Delta Y / \Delta Z = 2$ \citep{Casagrande_2007}, corresponds to $Y = 0.283$ and $Z = 0.0175$.

To obtain a consistent set of magnitudes for both the Dartmouth and Pisa models, we transform the theoretical isochrones from the ($\log T_{eff}$, $\log L/L_\odot$) plane into the color-magnitude diagram by means of our own calculation of photometric band. To do this we use synthetic spectra obtained from detailed atmospheric models following the method described in Section \ref{synth_phot}. 

\subsection{Atmospheric models}
To calculate synthetic photometry we use synthetic spectra provided by BT-Settl atmospheric models \citep{Allard_etal_2013}.
The BT-Settl models take into account gravitational dust settling in atmospheres of objects at temperatures below $\sim$2600 K, following the approach described in \citet{Rossow_1978}.
Opacities are introduced line-by-line to account for the effect of molecular absorbers, as described in \citet{Allard_etal_2003}.
The synthetic spectra are calculated using the radiative transfer model atmosphere code PHOENIX that implements static and radial (1D) approximations \citep{Allard_etal_2001}.
We apply the latest publicly available release of BT-Settl (last update in March 2015) that uses the \citet{Caffau_etal_2011} solar abundances. For effective temperatures $T_{eff} < 7000$ K we use the BT-Settl 2010 version that exploits solar abundances of \citet{Asplund_etal_2009}.

\subsection{Synthetic photometry with TA-DA}\label{synth_phot}
TA-DA stands for the Tool for Astrophysical Data Analysis \citep{Da_Rio_Roberrto_2012}. TA-DA is an interactive software allowing to analyze stellar photometric data in comparison with theoretical models and to derive stellar parameters using multi-band photometry.
TA-DA is able to interpolate stellar interior models and to produce synthetic photometry by converting stellar parameters into photometric magnitudes in given filters using grids of synthetic spectra. The conversion is done in the standard way, by integrating the synthetic spectra over the filter bandwidths and normalizing onto a spectrum of Vega. Therefore, TA-DA is well-suitable for combining various stellar interior and atmospheric models to derive photometric magnitudes for different evolutionary tracks and isochrones. 
We use stellar interior models and synthetic spectra from grids of atmospheric models to obtain various isochrones for the Hyades with TA-DA.

\section{Hyades single-star sequence}
In this section, we give the Hyades single-star sequence as an example for testing theoretical models. Additionally, we determine stellar parameters of the sequence members by comparing photometric observations to the calculated model isochrones. Absolute $JHKs$ magnitudes for the stars on the single star sequences (see Table 2) are derived from the distance estimates in \cite{Roeser_etal_2011} and the available near-infrared (NIR) photometry. For all stars we assume the NIR foreground extinction to be negligible.

\subsection{Isochrone comparison}\label{4_1}
We compute theoretical isochrones for combinations of BT-Settl 2010 atmospheric models with PISA and DARTMOUTH interior models for the age of 630 Myr, which is close to the 625$\pm$50 Myr  estimate of \cite{Perryman_etal_1998}. We place the obtained isochrones on $J $ vs. $J - K_s$, $J$ vs. $J - H$ and $K_s$ vs. $H - K_s$ color-magnitude diagrams together with the observed single-star sequence of the Hyades. As can be seen, both PISA and DARTMOUTH tracks predict the observed sequence reasonably well, even the  behavior of the "knee-shaped" part, around $\sim0.6M_{\odot}$ which had proven to be problematic as previously has been shown by \cite{Roeser_etal_2011} for older generation evolutionary models. The PISA models are available with ($\beta =0.2$) and without convective core overshooting. Both sets of models are in good agreement with the presented data, the only difference being the cluster inferred age. As expected the inclusion of convective core overshooting leads to older ages. Adopting $\beta=0.2$ the best fitting isochrone provides an age of 630 Myr, whereas isochrones without core overshooting give 550 Myr.

The observed sequence reveal a larger scatter towards lower masses. This is because, due to the faintness of the objects, both photometric and kinematic measurements have larger errors. Additionally, a small ($\sim$0.05 mag) systematic offset can be noted on $J$ vs. $J - K_s$  and $K_s$ vs $H - K_s$ diagrams -- the observed sequence shifted to redder colors from the theoretical predictions. 

\begin{figure*}[h]
	\centering
\begin{subfigure}[b]{0.45\textwidth}
		\includegraphics[width=9cm]{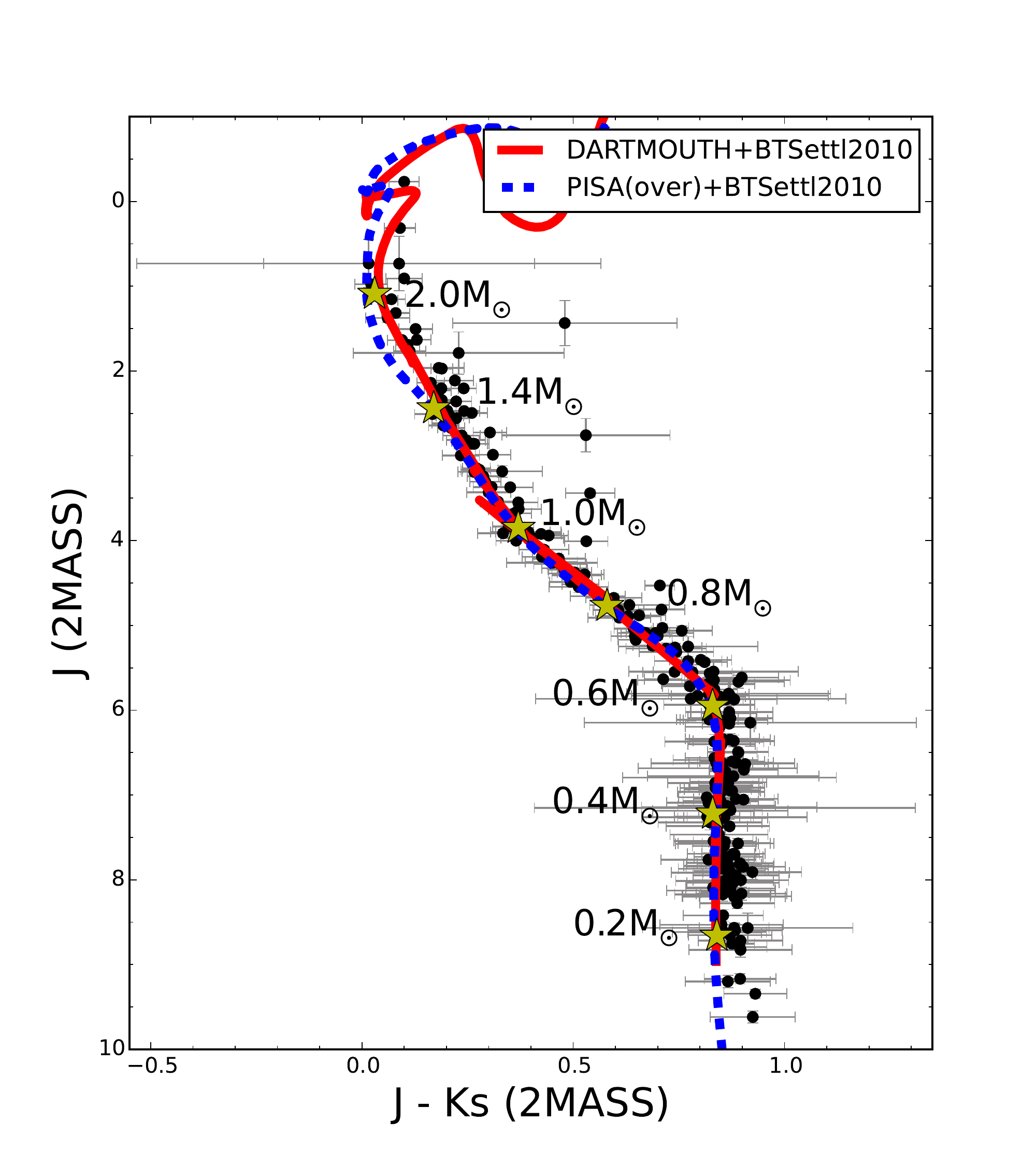}
		
	\end{subfigure}
	\begin{subfigure}[b]{0.45\textwidth}
		\includegraphics[width=9cm]{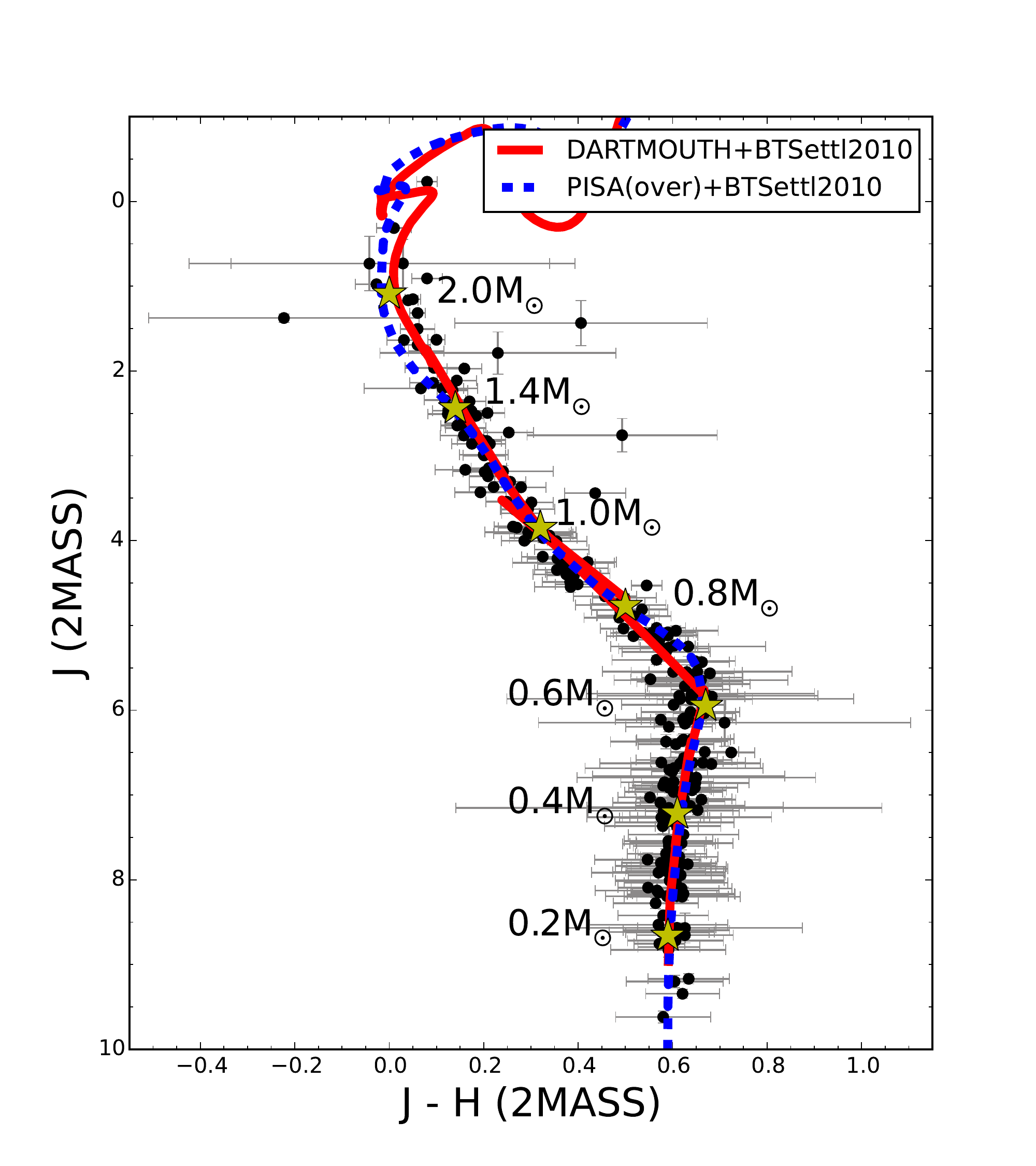}
		
	\end{subfigure}
	\label{iso_1}
\end{figure*}

\begin{SCfigure*}
\centering
\includegraphics[width=9cm]{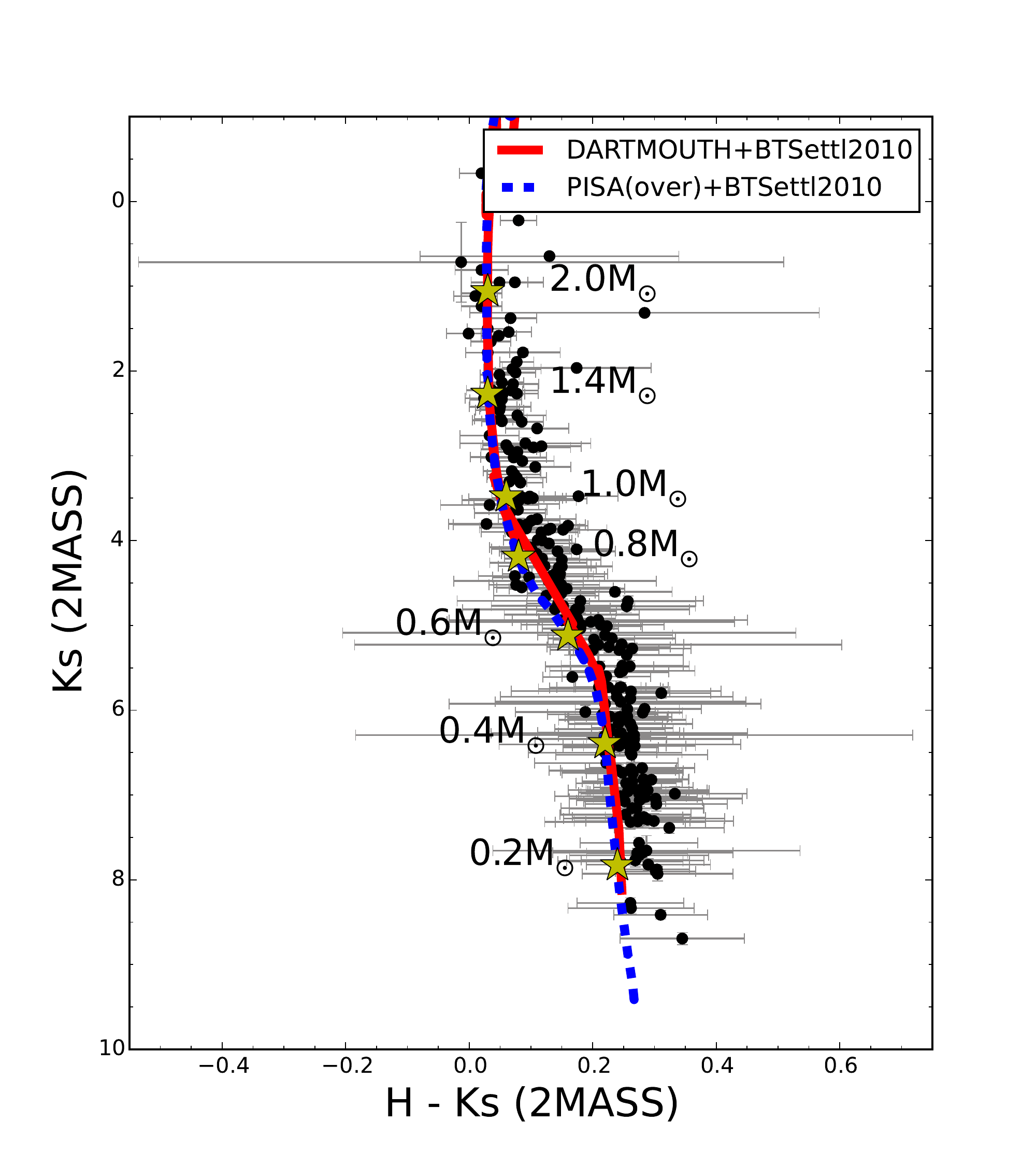}
\caption{$J$ vs $J - K_s$ (top left panel), $J$ vs. $J - H$ (top right panel) and $K_s$ vs. $H - K_s$ (bottom right panel) color-magnitude diagrams. The Hyades observed single-star sequence is over-plotted with theoretical isochrones (630 Myr) from different evolutionary models - PISA (dashed blue line) and DARTMOUTH (solid red line). Synthetic photometry for isochrones are calculated in TA-DA using BT-Settl 2010 synthetic spectra.}

\end{SCfigure*}

\subsection{Physical parameters of single stars with TA-DA}
TA-DA (see Section \ref{synth_phot}) includes an option that gives theoretical predictions for stellar parameters based on a comparison of synthetic photometry with observed photometric magnitudes. The dimension of the observational space (number of colors or magnitudes) must be greater or equal to the number of free parameters given by stellar interior models. The TA-DA parameter filter is described in \citet{Da_Rio_Roberrto_2012} and performs a multi-band least-square fit.
TA-DA also allows to estimate uncertainties for derived stellar parameters using a Monte Carlo simulation, in which the photometry is displaced according to photometric errors.

Using TA-DA we derive theoretical stellar parameters for 250 members of the Hyades single-star sequence (short electronic version of table Table 2) using BT-Settl2010+PISA isochrones for [Fe/H]=+0.14.

\subsection{Spectral Energy distribution from X-ray to MIR}
For all stars of the single star sequence, we compiled the available literature data from the X-rays to mid-infrared ROSAT X-ray \citep{Stern_etal_1995}, GALEX UV \citep{Martin_etal_2005}, APASS $gri$ \citep{Henden_etal_2015}, 2MASS $JHK_s$, and WISE $W1$ to $W4$), and computed absolute fluxes and magnitudes based on the parallaxes from \citet{Roeser_etal_2011}. The values are provided in the long electronic version of Table 2.

\section{Summary and future prospects}
Using previously published high angular resolution and spectroscopic studies and our own Lucky Imaging AstraLux Norte observations, we compiled a "single-star sequence" for the Hyades, which at an average distance of 45 pc is the open cluster closest to the Sun. In total, we identified 250 members of the cluster that do not show signatures of a companion outside 2--7 AU projected separation (depending on the distance to each individual member and the survey).

Comparison of the near-infrared properties of the single-star sequence with theoretical isochrones based on PISA and DARTMOUTH stellar interior models and BT-Settl 2010 atmospheric models shows an overall good agreement for the mass range 0.13 to 2.6 M$\odot$. The only disagreement between models and observations is that for masses below 0.6 Msun the observed $J - K_s$ and $H - K_s$ colors are systematically redder than the isochrones by 0.05 mag. But despite this, the isochrones calculated using most recent interior and atmospheric models, show a significantly better agreement with observations than isochrones produced by older generation models. The improvement can be particularly seen around 0.6--0.8 M$\odot$, where new isochrones are able to reproduce the "knee-shaped" part of the observed sequence.

We also estimate physical parameters (mass,effective temperature and surface gravity) for the Hyades single-star sequence members based on theoretical predictions of combined PISA interior and BT-Settl 2010 atmospheric models.

The resulting "single-star sequence" can be used for testing various theoretical models and also for selecting candidates to search for close, previously unresolved binaries using spectroscopy and high-resolution imaging instruments. In the near future, results from the Gaia survey will provide us with kinematic, photometric and spectroscopic information that will help to reduce the scatter at the lowest-mass part of the observed sequence, which will enable us to test theoretical models with stronger confidence.

\begin{acknowledgements} 
E.S. and S.R. acknowledge support by the
 Collaborative Research Center "The Milky Way System"
 (SFB 881, subproject B5)
 of the German Research Foundation (DFG). This publication makes use of data products from the Two Micron All Sky Survey, which is a joint project of the University of Massachusetts and the Infrared Processing and Analysis Center/California Institute of Technology, funded by the National Aeronautics and Space Administration and the National Science Foundation. This research has made use of
the SIMBAD database, operated at CDS, Strasbourg, France. We acknowledge with thanks the variable star observations from the AAVSO International Database contributed by observers worldwide and used in this research. The authors are grateful to Katie Morzinski for comments about her work on the Hyades.
\end{acknowledgements}

\onllongtab{
\begin{longtab}
\begin{landscape}
\begin{longtable}{rcccccccccccc}
\caption{Absolute $JHK_s$(2MASS) photometry and physical parameters of the Hyades single-star sequence members. The Resolution column correspond to values outside which no companion was detected. Objects observed only by the spectroscopic survey of \cite{Mermilliod_etal_2009} have no Resolution value. Hyades IDs are from \cite{Roeser_etal_2011}.}\\
\hline\hline
Hyades ID& parallax [mas] & parallax error [mas] & M$_{J}$ & dM$_{J}$& M$_{H}$ & dM$_{H}$& M$_{K}$ & dM$_{K}$& T$_{eff}$ [K] & $\log(g)$ (cgs)& Mass [M$\odot$]&  Resolution [AU] \\
\hline
\endfirsthead
\caption{continued.}\\
\hline\hline
Hyades ID& parallax [mas] & parallax error [mas] & M$_{J}$ & dM$_{J}$& M$_{H}$ & dM$_{H}$& M$_{K}$ & dM$_{K}$& T$_{eff}$ [K] & $\log(g)$ (cgs)& Mass [M$\odot$]&  Resolution [AU] \\
\hline
\endhead
\hline
\endfoot
3	&	28.1	&	1.08	&	7.32	&	0.09	&	6.72	&	0.09	&	6.50	&	0.09	&	3500	&	4.90	&	0.38	&	3.91	\\
6	&	31.9	&	0.1	&	3.64	&	0.02	&	3.37	&	0.02	&	3.30	&	0.02	&	5883	&	4.48	&	1.06	&	3.45	\\
9	&	22.36	&	0.41	&	7.03	&	0.05	&	6.48	&	0.05	&	6.21	&	0.05	&	3569	&	4.87	&	0.43	&	4.92	\\
10	&	23.23	&	0.41	&	6.62	&	0.05	&	6.04	&	0.04	&	5.78	&	0.04	&	3701	&	4.80	&	0.49	&	4.74	\\
11	&	23.41	&	0.37	&	5.84	&	0.04	&	5.16	&	0.06	&	4.99	&	0.04	&	4089	&	4.70	&	0.62	&	4.70	\\
13	&	34	&	0.36	&	7.37	&	0.03	&	6.76	&	0.03	&	6.50	&	0.03	&	3495	&	4.91	&	0.38	&	3.24	\\
16	&	19.73	&	0.13	&	4.55	&	0.07	&	4.16	&	0.04	&	4.03	&	0.02	&	5076	&	4.60	&	0.84	&	5.58	\\
21	&	19.46	&	0.4	&	6.11	&	0.05	&	5.49	&	0.05	&	5.29	&	0.05	&	3924	&	4.73	&	0.57	&	5.65	\\
22	&	23.5	&	0.4	&	7.70	&	0.04	&	7.12	&	0.04	&	6.82	&	0.04	&	3428	&	4.95	&	0.33	&	4.68	\\
25	&	26.37	&	0.16	&	2.99	&	0.03	&	2.79	&	0.04	&	2.68	&	0.03	&	6412	&	4.35	&	1.24	&	4.17	\\
27	&	24.58	&	0.13	&	4.34	&	0.03	&	3.95	&	0.02	&	3.86	&	0.03	&	5287	&	4.57	&	0.89	&	4.48	\\
28	&	30.89	&	0.04	&	2.76	&	0.20	&	2.26	&	0.04	&	2.23	&	0.02	&	6933	&	4.25	&	1.42	&	3.56	\\
30	&	30.26	&	0.42	&	7.82	&	0.04	&	7.18	&	0.03	&	6.92	&	0.04	&	3412	&	4.97	&	0.31	&	3.64	\\
34	&	19.24	&	0.45	&	7.26	&	0.06	&	6.67	&	0.06	&	6.44	&	0.05	&	3514	&	4.90	&	0.39	&	5.72	\\
37	&	27.85	&	0.44	&	8.15	&	0.04	&	7.58	&	0.04	&	7.29	&	0.04	&	3340	&	5.00	&	0.26	&	3.95	\\
39	&	21.35	&	0.15	&	6.62	&	0.03	&	5.98	&	0.03	&	5.73	&	0.03	&	3713	&	4.80	&	0.50	&	5.15	\\
40	&	25.36	&	0.14	&	3.85	&	0.03	&	3.58	&	0.03	&	3.48	&	0.03	&	5713	&	4.51	&	1.00	&	4.34	\\
43	&	29.08	&	0.43	&	6.63	&	0.04	&	5.95	&	0.04	&	5.73	&	0.04	&	3715	&	4.80	&	0.50	&	3.78	\\
49	&	22.27	&	0.44	&	6.04	&	0.05	&	5.37	&	0.05	&	5.17	&	0.05	&	3977	&	4.72	&	0.59	&	4.94	\\
50	&	30.63	&	0.55	&	5.66	&	0.05	&	5.00	&	0.05	&	4.83	&	0.04	&	4198	&	4.68	&	0.65	&	3.59	\\
51	&	16.64	&	0.18	&	3.94	&	0.03	&	3.60	&	0.03	&	3.50	&	0.04	&	5662	&	4.52	&	0.99	&	6.61	\\
52	&	32.16	&	0.2	&	5.08	&	0.03	&	4.49	&	0.05	&	4.42	&	0.03	&	4615	&	4.64	&	0.74	&	3.42	\\
53	&	31.71	&	0.45	&	5.75	&	0.04	&	5.10	&	0.05	&	4.92	&	0.04	&	4136	&	4.69	&	0.63	&	3.47	\\
57	&	26.66	&	0.5	&	7.19	&	0.05	&	6.58	&	0.05	&	6.33	&	0.05	&	3536	&	4.88	&	0.41	&	4.13	\\
61	&	19.7	&	0.11	&	2.83	&	0.02	&	2.62	&	0.02	&	2.58	&	0.02	&	6533	&	4.33	&	1.28	&	5.58	\\
65	&	27.08	&	0.21	&	6.72	&	0.03	&	6.12	&	0.03	&	5.86	&	0.03	&	3671	&	4.82	&	0.48	&	4.06	\\
66	&	25.81	&	0.48	&	6.16	&	0.05	&	5.53	&	0.05	&	5.29	&	0.04	&	3908	&	4.74	&	0.57	&	4.26	\\
70	&	28.1	&	0.5	&	7.81	&	0.04	&	7.21	&	0.04	&	6.96	&	0.04	&	3409	&	4.97	&	0.31	&	3.91	\\
71	&	18.85	&	0.2	&	5.12	&	0.04	&	4.53	&	0.05	&	4.43	&	0.03	&	4589	&	4.64	&	0.73	&	5.84	\\
73	&	24.14	&	0.48	&	6.95	&	0.05	&	6.33	&	0.05	&	6.08	&	0.05	&	3602	&	4.85	&	0.45	&	4.56	\\
74	&	24.81	&	1.5	&	6.68	&	0.13	&	6.08	&	0.13	&	5.84	&	0.13	&	3681	&	4.81	&	0.49	&	4.43	\\
77	&	27.2	&	0.48	&	7.94	&	0.04	&	7.33	&	0.05	&	7.08	&	0.05	&	3385	&	4.98	&	0.29	&	4.04	\\
78	&	21.82	&	0.49	&	7.18	&	0.05	&	6.53	&	0.05	&	6.31	&	0.05	&	3543	&	4.88	&	0.41	&	5.04	\\
79	&	18.1	&	0.56	&	7.09	&	0.07	&	6.52	&	0.07	&	6.27	&	0.07	&	3555	&	4.87	&	0.42	&	6.08	\\
82	&	26.12	&	0.49	&	6.89	&	0.05	&	6.31	&	0.05	&	6.03	&	0.04	&	3616	&	4.84	&	0.45	&	4.21	\\
83	&	21.12	&	0.22	&	4.40	&	0.03	&	4.02	&	0.06	&	3.87	&	0.03	&	5243	&	4.58	&	0.88	&	5.21	\\
85	&	24.25	&	0.43	&	5.03	&	0.04	&	4.46	&	0.04	&	4.32	&	0.04	&	4678	&	4.63	&	0.75	&	2.47	\\
86	&	22.17	&	0.25	&	5.43	&	0.04	&	4.77	&	0.04	&	4.62	&	0.04	&	4362	&	4.66	&	0.68	&	4.96	\\
88	&	23.83	&	0.06	&	2.14	&	0.03	&	2.05	&	0.04	&	1.98	&	0.02	&	7311	&	4.22	&	1.54	&	4.62	\\
89	&	14.21	&	0.24	&	4.25	&	0.04	&	3.83	&	0.04	&	3.80	&	0.04	&	5381	&	4.56	&	0.91	&	7.74	\\
93	&	22.31	&	0.54	&	6.70	&	0.06	&	6.11	&	0.06	&	5.80	&	0.06	&	3681	&	4.81	&	0.48	&	4.93	\\
94	&	21.26	&	0.36	&	4.82	&	0.04	&	4.33	&	0.04	&	4.21	&	0.04	&	4847	&	4.62	&	0.79	&	5.17	\\
95	&	24.69	&	0.23	&	4.38	&	0.03	&	4.00	&	0.03	&	3.87	&	0.03	&	5252	&	4.58	&	0.88	&	2.43	\\
97	&	17.49	&	1.43	&	6.79	&	0.18	&	6.14	&	0.18	&	5.92	&	0.18	&	3653	&	4.82	&	0.47	&	6.29	\\
99	&	21.91	&	0.52	&	6.93	&	0.06	&	6.33	&	0.06	&	6.09	&	0.06	&	3602	&	4.85	&	0.45	&	5.02	\\
100	&	20.96	&	0.51	&	6.40	&	0.06	&	5.79	&	0.06	&	5.55	&	0.06	&	3792	&	4.77	&	0.53	&	5.25	\\
101	&	22	&	0.49	&	7.69	&	0.05	&	7.10	&	0.05	&	6.81	&	0.05	&	3430	&	4.95	&	0.33	&	5.00	\\
104	&	22.68	&	0.61	&	8.62	&	0.06	&	8.03	&	0.06	&	7.77	&	0.06	&	3254	&	5.05	&	0.21	&	2.20	\\
105	&	21.31	&	0.49	&	7.27	&	0.05	&	6.66	&	0.05	&	6.42	&	0.05	&	3516	&	4.90	&	0.39	&	5.16	\\
106	&	13.46	&	0.2	&	3.37	&	0.04	&	3.09	&	0.04	&	3.02	&	0.04	&	6116	&	4.42	&	1.13	&	8.17	\\
107	&	19.84	&	0.49	&	6.84	&	0.06	&	6.24	&	0.06	&	5.99	&	0.06	&	3631	&	4.83	&	0.46	&	5.54	\\
108	&	23.2	&	1.49	&	5.54	&	0.14	&	4.89	&	0.14	&	4.71	&	0.14	&	4278	&	4.67	&	0.67	&	4.74	\\
109	&	25.42	&	0.51	&	6.56	&	0.05	&	5.94	&	0.05	&	5.73	&	0.05	&	3726	&	4.79	&	0.50	&	4.33	\\
112	&	22.17	&	0.51	&	6.60	&	0.05	&	5.97	&	0.05	&	5.73	&	0.05	&	3716	&	4.80	&	0.50	&	4.96	\\
114	&	24.07	&	0.22	&	2.47	&	0.03	&	2.30	&	0.04	&	2.23	&	0.03	&	6875	&	4.26	&	1.40	&	4.57	\\
116	&	21.03	&	0.53	&	8.19	&	0.06	&	7.58	&	0.06	&	7.31	&	0.06	&	3336	&	5.01	&	0.26	&	2.38	\\
118	&	22.24	&	0.05	&	1.97	&	0.02	&	1.81	&	0.03	&	1.78	&	0.02	&	7600	&	4.19	&	1.64	&	4.95	\\
120	&	21.97	&	0.21	&	4.75	&	0.03	&	4.26	&	0.05	&	4.16	&	0.03	&	4920	&	4.61	&	0.81	&	5.01	\\
123	&	23.25	&	0.54	&	7.60	&	0.05	&	7.01	&	0.05	&	6.75	&	0.05	&	3443	&	4.94	&	0.34	&	2.15	\\
124	&	20.56	&	0.46	&	7.55	&	0.05	&	6.96	&	0.05	&	6.69	&	0.05	&	3455	&	4.93	&	0.35	&	5.35	\\
128	&	19.97	&	0.59	&	8.42	&	0.07	&	7.84	&	0.07	&	7.56	&	0.07	&	3289	&	5.03	&	0.23	&	2.50	\\
133	&	21.5	&	0.77	&	5.55	&	0.08	&	4.92	&	0.08	&	4.77	&	0.08	&	4256	&	4.68	&	0.66	&	2.79	\\
134	&	22.28	&	0.24	&	4.49	&	0.04	&	4.10	&	0.04	&	3.99	&	0.03	&	5135	&	4.59	&	0.85	&	2.69	\\
136	&	25.16	&	0.54	&	6.49	&	0.05	&	5.82	&	0.05	&	5.60	&	0.05	&	3770	&	4.78	&	0.52	&	4.37	\\
137	&	21.04	&	0.19	&	3.37	&	0.03	&	3.15	&	0.04	&	3.06	&	0.03	&	6095	&	4.43	&	1.12	&	5.23	\\
138	&	20.45	&	0.53	&	7.89	&	0.06	&	7.31	&	0.06	&	7.02	&	0.06	&	3393	&	4.98	&	0.30	&	2.44	\\
139	&	20.11	&	0.21	&	6.50	&	0.03	&	5.78	&	0.04	&	5.61	&	0.03	&	3774	&	4.78	&	0.52	&	5.47	\\
140	&	22.47	&	0.35	&	5.12	&	0.04	&	4.57	&	0.06	&	4.42	&	0.04	&	4581	&	4.64	&	0.73	&	4.90	\\
141	&	29.42	&	0.56	&	8.79	&	0.05	&	8.20	&	0.05	&	7.90	&	0.05	&	3227	&	5.07	&	0.19	&	1.70	\\
145	&	23.76	&	1.37	&	5.65	&	0.13	&	4.99	&	0.13	&	4.81	&	0.13	&	4208	&	4.68	&	0.65	&	2.53	\\
146	&	17.83	&	0.2	&	3.44	&	0.04	&	3.00	&	0.05	&	2.90	&	0.04	&	6159	&	4.41	&	1.15	&	6.17	\\
147	&	20.92	&	0.47	&	5.42	&	0.06	&	4.77	&	0.06	&	4.65	&	0.06	&	4357	&	4.67	&	0.68	&	2.87	\\
148	&	21.63	&	0.24	&	4.52	&	0.03	&	4.12	&	0.03	&	4.00	&	0.03	&	5117	&	4.59	&	0.85	&	2.77	\\
152	&	28.29	&	0.29	&	5.17	&	0.03	&	4.60	&	0.03	&	4.52	&	0.03	&	4523	&	4.65	&	0.72	&	3.89	\\
153	&	28.3	&	0.62	&	9.34	&	0.05	&	8.72	&	0.06	&	8.41	&	0.05	&	3128	&	5.13	&	0.15	&	1.77	\\
155	&	22.51	&	0.09	&	2.56	&	0.02	&	2.39	&	0.02	&	2.33	&	0.02	&	6778	&	4.28	&	1.37	&	4.89	\\
157	&	21.6	&	0.54	&	7.27	&	0.06	&	6.69	&	0.06	&	6.42	&	0.06	&	3513	&	4.90	&	0.39	&	5.09	\\
159	&	26.83	&	0.11	&	2.21	&	0.02	&	2.14	&	0.12	&	1.96	&	0.02	&	7229	&	4.22	&	1.52	&	4.10	\\
165	&	27.65	&	0.56	&	6.94	&	0.05	&	6.30	&	0.05	&	6.08	&	0.05	&	3606	&	4.85	&	0.45	&	3.98	\\
167	&	25.57	&	0.66	&	9.17	&	0.06	&	8.53	&	0.06	&	8.27	&	0.06	&	3163	&	5.11	&	0.16	&	1.96	\\
168	&	23.12	&	0.81	&	7.76	&	0.08	&	7.22	&	0.08	&	6.94	&	0.08	&	3412	&	4.97	&	0.31	&	4.76	\\
174	&	21.73	&	0.24	&	3.54	&	0.03	&	3.29	&	0.03	&	3.22	&	0.03	&	5959	&	4.46	&	1.08	&	2.76	\\
175	&	20.48	&	0.58	&	6.85	&	0.06	&	6.27	&	0.07	&	5.98	&	0.06	&	3628	&	4.84	&	0.46	&	5.37	\\
176	&	17.87	&	0.09	&	2.47	&	0.02	&	2.34	&	0.03	&	2.27	&	0.02	&	6849	&	4.26	&	1.39	&	6.16	\\
177	&	22.91	&	0.17	&	3.68	&	0.03	&	3.40	&	0.03	&	3.32	&	0.03	&	5861	&	4.48	&	1.05	&	2.62	\\
178	&	22.19	&	0.13	&	2.86	&	0.02	&	2.65	&	0.02	&	2.59	&	0.02	&	6512	&	4.33	&	1.27	&	4.96	\\
179	&	18.05	&	0.61	&	8.16	&	0.08	&	7.54	&	0.08	&	7.27	&	0.08	&	3343	&	5.00	&	0.26	&	2.77	\\
183	&	22	&	0.55	&	5.62	&	0.06	&	4.97	&	0.09	&	4.72	&	0.06	&	4243	&	4.68	&	0.66	&	5.00	\\
184	&	22.33	&	0.77	&	5.69	&	0.08	&	5.04	&	0.08	&	4.87	&	0.08	&	4173	&	4.69	&	0.64	&	2.69	\\
187	&	39.21	&	0.58	&	8.75	&	0.04	&	8.18	&	0.04	&	7.88	&	0.04	&	3231	&	5.07	&	0.19	&	2.81	\\
190	&	21.8	&	0.11	&	2.21	&	0.04	&	2.09	&	0.03	&	2.02	&	0.02	&	7223	&	4.22	&	1.51	&	5.05	\\
192	&	22.55	&	0.59	&	7.57	&	0.06	&	6.96	&	0.06	&	6.68	&	0.06	&	3454	&	4.93	&	0.35	&	4.88	\\
193	&	21.57	&	0.56	&	6.92	&	0.06	&	6.32	&	0.06	&	6.08	&	0.06	&	3607	&	4.85	&	0.45	&	5.10	\\
194	&	22.4	&	0.7	&	9.20	&	0.07	&	8.60	&	0.07	&	8.33	&	0.07	&	3152	&	5.12	&	0.15	&	2.23	\\
195	&	22.58	&	0.4	&	5.83	&	0.04	&	5.22	&	0.04	&	5.04	&	0.04	&	4070	&	4.70	&	0.61	&	2.66	\\
198	&	19.85	&	0.23	&	4.19	&	0.03	&	3.86	&	0.03	&	3.76	&	0.03	&	5408	&	4.56	&	0.92	&	3.02	\\
201	&	21.27	&	0.59	&	7.07	&	0.06	&	6.45	&	0.06	&	6.22	&	0.06	&	3566	&	4.87	&	0.43	&	5.17	\\
202	&	21.31	&	0.24	&	5.24	&	0.03	&	4.64	&	0.03	&	4.55	&	0.03	&	4477	&	4.65	&	0.71	&	2.82	\\
203	&	19.89	&	0.64	&	6.96	&	0.07	&	6.36	&	0.07	&	6.11	&	0.07	&	3594	&	4.85	&	0.44	&	5.53	\\
204	&	23.78	&	0.57	&	7.80	&	0.06	&	7.23	&	0.06	&	6.95	&	0.06	&	3409	&	4.97	&	0.31	&	4.63	\\
210	&	19.06	&	0.64	&	7.27	&	0.08	&	6.68	&	0.08	&	6.42	&	0.07	&	3514	&	4.90	&	0.39	&	5.77	\\
213	&	21.17	&	0.58	&	8.28	&	0.06	&	7.71	&	0.06	&	7.39	&	0.06	&	3317	&	5.02	&	0.25	&	2.36	\\
216	&	22.67	&	0.43	&	4.66	&	0.05	&	4.20	&	0.05	&	4.10	&	0.04	&	4993	&	4.61	&	0.82	&	2.65	\\
223	&	21.79	&	0.65	&	5.81	&	0.07	&	5.17	&	0.07	&	4.99	&	0.07	&	4092	&	4.70	&	0.62	&	5.05	\\
225	&	22.03	&	0.29	&	4.00	&	0.03	&	3.72	&	0.03	&	3.64	&	0.03	&	5569	&	4.54	&	0.96	&	4.99	\\
226	&	24.01	&	0.59	&	7.81	&	0.06	&	7.19	&	0.06	&	6.91	&	0.06	&	3413	&	4.97	&	0.31	&	4.58	\\
227	&	21.89	&	0.18	&	3.84	&	0.03	&	3.57	&	0.03	&	3.49	&	0.02	&	5713	&	4.51	&	1.00	&	2.74	\\
228	&	20.08	&	0.56	&	6.62	&	0.06	&	5.96	&	0.06	&	5.74	&	0.06	&	3715	&	4.80	&	0.50	&	5.48	\\
229	&	22.33	&	0.08	&	-1.29	&	0.25	&	-1.76	&	0.21	&	-1.74	&	0.20	&	6062	&	3.24	&	2.30	&	4.93	\\
233	&	22.39	&	0.15	&	3.15	&	0.02	&	2.93	&	0.03	&	2.87	&	0.02	&	6265	&	4.39	&	1.18	&	4.91	\\
234	&	23.82	&	0.11	&	2.23	&	0.02	&	2.09	&	0.03	&	2.04	&	0.02	&	7173	&	4.23	&	1.50	&	4.62	\\
237	&	21.14	&	0.09	&	1.38	&	0.05	&	1.60	&	0.28	&	1.31	&	0.02	&	8140	&	4.08	&	1.90	&	5.20	\\
242	&	21.63	&	0.74	&	8.04	&	0.08	&	7.43	&	0.08	&	7.16	&	0.08	&	3367	&	4.99	&	0.28	&	2.31	\\
244	&	20.5	&	0.11	&	2.53	&	0.02	&	2.34	&	0.02	&	2.32	&	0.02	&	6809	&	4.27	&	1.38	&	5.37	\\
245	&	21.61	&	0.13	&	2.62	&	0.02	&	2.47	&	0.02	&	2.42	&	0.02	&	6703	&	4.29	&	1.34	&	5.09	\\
246	&	20.5	&	0.12	&	2.34	&	0.03	&	2.22	&	0.04	&	2.15	&	0.02	&	7021	&	4.25	&	1.45	&	5.37	\\
247	&	20.64	&	0.69	&	7.87	&	0.08	&	7.26	&	0.08	&	7.01	&	0.07	&	3399	&	4.98	&	0.30	&	2.42	\\
248	&	25.75	&	0.86	&	7.84	&	0.08	&	7.25	&	0.08	&	6.97	&	0.07	&	3404	&	4.97	&	0.31	&	1.94	\\
249	&	25.78	&	0.65	&	8.57	&	0.06	&	7.96	&	0.06	&	7.69	&	0.06	&	3267	&	5.05	&	0.21	&	1.94	\\
250	&	20.89	&	1.07	&	7.19	&	0.11	&	6.61	&	0.11	&	6.34	&	0.11	&	3533	&	4.89	&	0.41	&	5.27	\\
251	&	27.16	&	0.58	&	7.05	&	0.05	&	6.39	&	0.05	&	6.15	&	0.05	&	3579	&	4.86	&	0.43	&	4.05	\\
256	&	18.65	&	0.68	&	7.13	&	0.08	&	6.49	&	0.08	&	6.27	&	0.08	&	3554	&	4.87	&	0.42	&	2.68	\\
257	&	27.21	&	1.78	&	6.78	&	0.14	&	6.14	&	0.14	&	5.90	&	0.14	&	3658	&	4.82	&	0.47	&	2.21	\\
258	&	20.78	&	0.65	&	7.05	&	0.07	&	6.42	&	0.07	&	6.16	&	0.07	&	3577	&	4.86	&	0.43	&	5.29	\\
261	&	21.1	&	0.15	&	-1.28	&	0.25	&	-1.80	&	0.20	&	-1.74	&	0.23	&	6062	&	3.24	&	2.30	&	5.21	\\
265	&	21.93	&	0.7	&	7.95	&	0.07	&	7.34	&	0.07	&	7.06	&	0.07	&	3385	&	4.98	&	0.29	&	2.28	\\
269	&	20.85	&	0.13	&	1.64	&	0.02	&	1.60	&	0.03	&	1.54	&	0.02	&	7947	&	4.15	&	1.78	&	5.28	\\
271	&	18.69	&	0.14	&	2.64	&	0.03	&	2.50	&	0.02	&	2.45	&	0.02	&	6673	&	4.30	&	1.33	&	5.89	\\
272	&	20.9	&	0.39	&	4.21	&	0.04	&	3.86	&	0.05	&	3.75	&	0.04	&	5409	&	4.56	&	0.92	&	2.87	\\
273	&	23.44	&	0.68	&	7.54	&	0.07	&	6.95	&	0.07	&	6.71	&	0.07	&	3454	&	4.93	&	0.35	&	4.69	\\
274	&	23.11	&	0.68	&	6.11	&	0.07	&	5.54	&	0.07	&	5.27	&	0.07	&	3918	&	4.74	&	0.57	&	4.76	\\
275	&	20.4	&	0.7	&	5.67	&	0.08	&	5.03	&	0.08	&	4.77	&	0.08	&	4202	&	4.68	&	0.65	&	5.39	\\
276	&	22.58	&	0.66	&	5.27	&	0.07	&	4.69	&	0.07	&	4.55	&	0.07	&	4447	&	4.66	&	0.70	&	2.66	\\
278	&	23.57	&	0.59	&	7.73	&	0.06	&	7.11	&	0.06	&	6.86	&	0.06	&	3425	&	4.96	&	0.32	&	4.67	\\
279	&	21.8	&	0.68	&	6.11	&	0.07	&	5.48	&	0.07	&	5.25	&	0.07	&	3931	&	4.73	&	0.57	&	2.75	\\
281	&	21.84	&	0.11	&	1.17	&	0.01	&	1.13	&	0.01	&	1.12	&	0.03	&	8185	&	4.02	&	1.98	&	5.04	\\
283	&	36.13	&	0.68	&	8.09	&	0.05	&	7.54	&	0.04	&	7.26	&	0.04	&	3348	&	5.00	&	0.27	&	3.04	\\
287	&	20.32	&	0.73	&	8.17	&	0.08	&	7.55	&	0.08	&	7.32	&	0.08	&	3339	&	5.01	&	0.26	&	2.46	\\
288	&	22.26	&	0.53	&	4.89	&	0.05	&	4.37	&	0.06	&	4.26	&	0.05	&	4785	&	4.63	&	0.78	&	2.70	\\
290	&	21.76	&	0.68	&	8.72	&	0.07	&	8.11	&	0.07	&	7.82	&	0.07	&	3241	&	5.06	&	0.20	&	2.30	\\
291	&	19.49	&	0.21	&	2.51	&	0.03	&	2.38	&	0.03	&	2.34	&	0.03	&	6795	&	4.27	&	1.37	&	5.64	\\
295	&	22.04	&	0.77	&	8.62	&	0.08	&	8.04	&	0.08	&	7.77	&	0.08	&	3254	&	5.05	&	0.20	&	2.27	\\
297	&	19.45	&	0.22	&	4.01	&	0.04	&	3.65	&	0.05	&	3.48	&	0.04	&	5639	&	4.53	&	0.98	&	5.66	\\
298	&	20.45	&	0.72	&	6.86	&	0.08	&	6.21	&	0.08	&	6.02	&	0.08	&	3629	&	4.84	&	0.46	&	5.38	\\
301	&	19.02	&	0.35	&	4.88	&	0.05	&	4.38	&	0.05	&	4.23	&	0.04	&	4801	&	4.62	&	0.78	&	5.78	\\
302	&	22.12	&	0.68	&	5.87	&	0.07	&	5.20	&	0.07	&	4.99	&	0.07	&	4071	&	4.70	&	0.61	&	4.97	\\
304	&	28.67	&	0.68	&	6.20	&	0.06	&	5.60	&	0.07	&	5.35	&	0.06	&	3883	&	4.74	&	0.56	&	3.84	\\
306	&	28.44	&	0.68	&	7.69	&	0.06	&	7.11	&	0.06	&	6.84	&	0.06	&	3428	&	4.95	&	0.33	&	3.87	\\
307	&	21.37	&	0.59	&	6.91	&	0.06	&	6.27	&	0.06	&	6.05	&	0.06	&	3615	&	4.84	&	0.45	&	5.15	\\
308	&	18.74	&	0.59	&	6.09	&	0.07	&	5.47	&	0.07	&	5.22	&	0.07	&	3940	&	4.73	&	0.58	&	5.87	\\
309	&	21.27	&	0.05	&	0.73	&	0.28	&	0.70	&	0.23	&	0.72	&	0.47	&	8091	&	3.88	&	2.14	&	5.17	\\
319	&	23.26	&	0.72	&	7.85	&	0.07	&	7.23	&	0.07	&	6.94	&	0.07	&	3406	&	4.97	&	0.31	&	2.15	\\
325	&	22.02	&	0.06	&	0.31	&	0.03	&	0.30	&	0.02	&	0.22	&	0.02	&	8547	&	3.75	&	2.27	&	5.00	\\
338	&	21.9	&	0.72	&	8.10	&	0.08	&	7.48	&	0.08	&	7.23	&	0.07	&	3354	&	5.00	&	0.27	&	2.28	\\
340	&	21.09	&	0.22	&	3.00	&	0.03	&	2.79	&	0.03	&	2.76	&	0.03	&	6380	&	4.36	&	1.22	&	5.22	\\
341	&	16.65	&	2.44	&	7.15	&	0.32	&	6.56	&	0.32	&	6.29	&	0.32	&	3544	&	4.88	&	0.41	&	3.00	\\
344	&	22.56	&	0.22	&	3.19	&	0.03	&	2.99	&	0.04	&	2.92	&	0.03	&	6224	&	4.40	&	1.17	&	4.88	\\
348	&	18.35	&	0.68	&	6.37	&	0.08	&	5.79	&	0.08	&	5.54	&	0.08	&	3800	&	4.77	&	0.53	&	5.99	\\
349	&	20.22	&	0.09	&	1.69	&	0.02	&	1.63	&	0.02	&	1.58	&	0.02	&	7900	&	4.15	&	1.75	&	5.44	\\
352	&	23.4	&	0.46	&	5.41	&	0.05	&	4.84	&	0.08	&	4.60	&	0.05	&	4367	&	4.66	&	0.69	&	4.70	\\
353	&	20.22	&	0.77	&	8.20	&	0.09	&	7.58	&	0.09	&	7.32	&	0.08	&	3336	&	5.01	&	0.26	&	2.47	\\
354	&	22.43	&	0.36	&	1.96	&	0.04	&	1.87	&	0.04	&	1.78	&	0.04	&	7586	&	4.19	&	1.63	&	0.00	\\
359	&	22.37	&	0.12	&	-0.23	&	0.02	&	-0.31	&	0.02	&	-0.33	&	0.03	&	7122	&	3.43	&	2.29	&	4.92	\\
361	&	21.74	&	0.06	&	1.32	&	0.01	&	1.26	&	0.01	&	1.24	&	0.03	&	8160	&	4.06	&	1.92	&	5.06	\\
367	&	21.46	&	0.58	&	6.34	&	0.06	&	5.70	&	0.06	&	5.49	&	0.06	&	3824	&	4.76	&	0.54	&	5.13	\\
368	&	21.17	&	0.71	&	5.94	&	0.08	&	5.33	&	0.08	&	5.11	&	0.08	&	4009	&	4.71	&	0.60	&	2.83	\\
370	&	21.86	&	0.16	&	2.73	&	0.03	&	2.47	&	0.04	&	2.42	&	0.03	&	6662	&	4.30	&	1.33	&	5.03	\\
371	&	22.03	&	0.17	&	2.76	&	0.03	&	2.60	&	0.04	&	2.52	&	0.03	&	6578	&	4.31	&	1.30	&	4.99	\\
373	&	20.78	&	0.3	&	4.81	&	0.04	&	4.28	&	0.04	&	4.10	&	0.04	&	4905	&	4.62	&	0.80	&	5.29	\\
374	&	22.41	&	0.37	&	4.28	&	0.04	&	3.90	&	0.04	&	3.81	&	0.04	&	5346	&	4.57	&	0.91	&	2.68	\\
378	&	23.15	&	0.68	&	6.35	&	0.07	&	5.72	&	0.07	&	5.48	&	0.07	&	3821	&	4.76	&	0.54	&	2.59	\\
379	&	21.96	&	0.3	&	4.33	&	0.04	&	3.92	&	0.04	&	3.85	&	0.04	&	5305	&	4.57	&	0.90	&	5.01	\\
386	&	21.81	&	0.06	&	1.15	&	0.01	&	1.10	&	0.01	&	1.08	&	0.03	&	8187	&	4.02	&	1.99	&	5.04	\\
389	&	22.41	&	0.08	&	1.63	&	0.01	&	1.53	&	0.01	&	1.50	&	0.03	&	7974	&	4.14	&	1.79	&	4.91	\\
397	&	21.63	&	0.17	&	2.81	&	0.04	&	2.62	&	0.03	&	2.57	&	0.03	&	6543	&	4.32	&	1.28	&	5.09	\\
399	&	18.76	&	1.18	&	7.26	&	0.14	&	6.65	&	0.14	&	6.40	&	0.14	&	3518	&	4.89	&	0.40	&	5.86	\\
400	&	21.13	&	0.81	&	8.83	&	0.09	&	8.23	&	0.09	&	7.93	&	0.09	&	3221	&	5.07	&	0.19	&	2.37	\\
401	&	21.27	&	0.44	&	3.90	&	0.05	&	3.60	&	0.08	&	3.52	&	0.05	&	5677	&	4.52	&	0.99	&	5.17	\\
403	&	21.94	&	0.77	&	8.00	&	0.08	&	7.41	&	0.08	&	7.11	&	0.08	&	3374	&	4.99	&	0.28	&	2.28	\\
404	&	25.23	&	0.77	&	8.65	&	0.07	&	8.03	&	0.07	&	7.78	&	0.07	&	3252	&	5.05	&	0.20	&	1.98	\\
405	&	19.34	&	0.21	&	3.17	&	0.03	&	3.00	&	0.06	&	2.89	&	0.03	&	6247	&	4.39	&	1.18	&	0.00	\\
409	&	20.72	&	1.09	&	5.25	&	0.12	&	4.62	&	0.12	&	4.48	&	0.12	&	4499	&	4.65	&	0.71	&	2.90	\\
410	&	24.01	&	0.15	&	2.49	&	0.03	&	2.29	&	0.02	&	2.23	&	0.03	&	6871	&	4.26	&	1.40	&	4.58	\\
412	&	23.17	&	0.71	&	7.58	&	0.07	&	6.99	&	0.07	&	6.74	&	0.07	&	3447	&	4.94	&	0.34	&	4.75	\\
415	&	17.01	&	0.75	&	7.92	&	0.10	&	7.35	&	0.10	&	7.04	&	0.10	&	3388	&	4.98	&	0.30	&	2.94	\\
417	&	20.12	&	0.26	&	5.09	&	0.04	&	4.55	&	0.05	&	4.42	&	0.04	&	4603	&	4.64	&	0.74	&	2.98	\\
419	&	22.11	&	0.76	&	7.57	&	0.08	&	6.95	&	0.08	&	6.71	&	0.08	&	3452	&	4.93	&	0.35	&	4.98	\\
424	&	18.31	&	0.33	&	5.06	&	0.06	&	4.46	&	0.07	&	4.31	&	0.04	&	4682	&	4.63	&	0.75	&	6.01	\\
428	&	23.27	&	0.4	&	5.26	&	0.04	&	4.67	&	0.04	&	4.52	&	0.05	&	4468	&	4.65	&	0.71	&	2.58	\\
429	&	20.4	&	0.23	&	4.39	&	0.03	&	3.97	&	0.04	&	3.90	&	0.03	&	5247	&	4.58	&	0.88	&	2.94	\\
431	&	24.39	&	0.71	&	7.84	&	0.07	&	7.26	&	0.07	&	6.98	&	0.07	&	3403	&	4.97	&	0.31	&	2.05	\\
432	&	22.02	&	0.15	&	2.47	&	0.02	&	2.33	&	0.02	&	2.27	&	0.03	&	6849	&	4.26	&	1.39	&	5.00	\\
433	&	22.27	&	0.21	&	0.91	&	0.02	&	0.83	&	0.02	&	0.81	&	0.04	&	8192	&	3.93	&	2.10	&	4.94	\\
434	&	22.35	&	0.79	&	8.60	&	0.08	&	7.99	&	0.08	&	7.71	&	0.08	&	3261	&	5.05	&	0.21	&	2.24	\\
439	&	18.6	&	0.29	&	4.41	&	0.04	&	4.02	&	0.04	&	3.90	&	0.04	&	5220	&	4.58	&	0.88	&	3.23	\\
440	&	21.74	&	0.29	&	3.43	&	0.03	&	3.24	&	0.04	&	3.13	&	0.04	&	6035	&	4.44	&	1.10	&	5.06	\\
455	&	19.82	&	0.25	&	4.35	&	0.03	&	3.99	&	0.03	&	3.86	&	0.03	&	5272	&	4.58	&	0.89	&	3.03	\\
456	&	24.07	&	0.09	&	2.11	&	0.04	&	1.97	&	0.02	&	1.89	&	0.02	&	7423	&	4.21	&	1.58	&	4.57	\\
459	&	23.02	&	0.72	&	6.59	&	0.07	&	5.96	&	0.07	&	5.75	&	0.07	&	3717	&	4.80	&	0.50	&	4.78	\\
460	&	20.7	&	0.15	&	2.36	&	0.03	&	2.19	&	0.02	&	2.14	&	0.03	&	7027	&	4.24	&	1.45	&	5.31	\\
461	&	52.1	&	0.23	&	4.53	&	0.02	&	3.98	&	0.02	&	3.82	&	0.02	&	5210	&	4.58	&	0.87	&	2.11	\\
465	&	19.77	&	0.81	&	8.19	&	0.09	&	7.61	&	0.09	&	7.31	&	0.09	&	3335	&	5.01	&	0.26	&	2.53	\\
466	&	19.88	&	0.81	&	8.13	&	0.09	&	7.56	&	0.09	&	7.28	&	0.09	&	3344	&	5.00	&	0.26	&	2.52	\\
475	&	21.17	&	0.77	&	8.01	&	0.08	&	7.42	&	0.08	&	7.15	&	0.08	&	3370	&	4.99	&	0.28	&	2.36	\\
478	&	23.47	&	0.15	&	3.31	&	0.03	&	3.05	&	0.02	&	3.01	&	0.03	&	6149	&	4.42	&	1.14	&	4.69	\\
480	&	23.06	&	0.24	&	3.63	&	0.04	&	3.33	&	0.04	&	3.26	&	0.03	&	5914	&	4.47	&	1.06	&	4.77	\\
486	&	23.12	&	0.51	&	4.91	&	0.05	&	4.42	&	0.05	&	4.30	&	0.05	&	4750	&	4.63	&	0.77	&	2.60	\\
487	&	20.36	&	0.75	&	7.91	&	0.08	&	7.32	&	0.08	&	6.98	&	0.08	&	3395	&	4.98	&	0.30	&	2.46	\\
489	&	23.2	&	0.3	&	5.57	&	0.04	&	4.89	&	0.03	&	4.74	&	0.04	&	4267	&	4.68	&	0.66	&	4.74	\\
491	&	16.68	&	0.28	&	4.11	&	0.04	&	3.74	&	0.04	&	3.68	&	0.04	&	5511	&	4.55	&	0.95	&	3.60	\\
496	&	14.05	&	1.13	&	8.57	&	0.18	&	7.94	&	0.18	&	7.66	&	0.18	&	3269	&	5.04	&	0.21	&	3.56	\\
497	&	39.4	&	0.36	&	5.87	&	0.03	&	5.23	&	0.05	&	5.01	&	0.03	&	4070	&	4.70	&	0.61	&	2.79	\\
498	&	15.48	&	0.16	&	3.55	&	0.03	&	3.25	&	0.03	&	3.18	&	0.03	&	5979	&	4.46	&	1.08	&	0.00	\\
501	&	25.01	&	0.79	&	9.62	&	0.07	&	9.04	&	0.07	&	8.69	&	0.07	&	3059	&	5.16	&	0.13	&	2.00	\\
502	&	14.51	&	0.19	&	3.92	&	0.04	&	3.59	&	0.05	&	3.50	&	0.03	&	5679	&	4.52	&	0.99	&	7.58	\\
518	&	23.18	&	0.09	&	1.79	&	0.25	&	1.56	&	0.03	&	1.56	&	0.02	&	7971	&	4.14	&	1.79	&	4.75	\\
525	&	15.71	&	0.14	&	0.98	&	0.03	&	1.00	&	0.04	&	0.95	&	0.03	&	8194	&	3.97	&	2.05	&	7.00	\\
528	&	22.31	&	0.61	&	5.31	&	0.06	&	4.73	&	0.07	&	4.57	&	0.06	&	4424	&	4.66	&	0.70	&	2.69	\\
539	&	22.29	&	0.72	&	6.02	&	0.07	&	5.38	&	0.07	&	5.15	&	0.07	&	3979	&	4.72	&	0.59	&	4.93	\\
542	&	25.89	&	0.46	&	5.72	&	0.04	&	5.09	&	0.07	&	4.94	&	0.05	&	4144	&	4.69	&	0.63	&	4.25	\\
544	&	19.04	&	0.46	&	5.09	&	0.07	&	4.53	&	0.06	&	4.39	&	0.06	&	4609	&	4.64	&	0.74	&	3.15	\\
545	&	21.26	&	0.28	&	5.55	&	0.04	&	4.95	&	0.03	&	4.81	&	0.03	&	4239	&	4.68	&	0.66	&	2.82	\\
549	&	27.6	&	0.27	&	5.63	&	0.03	&	5.08	&	0.03	&	4.92	&	0.03	&	4171	&	4.69	&	0.64	&	3.99	\\
558	&	22.28	&	0.75	&	4.26	&	0.08	&	3.89	&	0.08	&	3.81	&	0.08	&	5354	&	4.57	&	0.91	&	4.94	\\
566	&	20.89	&	0.18	&	3.24	&	0.03	&	3.03	&	0.03	&	2.95	&	0.02	&	6189	&	4.41	&	1.16	&	5.27	\\
571	&	21.3	&	0.35	&	5.13	&	0.04	&	4.61	&	0.04	&	4.48	&	0.04	&	4547	&	4.65	&	0.73	&	5.16	\\
576	&	20.3	&	0.16	&	2.86	&	0.03	&	2.68	&	0.03	&	2.60	&	0.02	&	6503	&	4.33	&	1.27	&	5.42	\\
578	&	23.24	&	0.29	&	5.63	&	0.04	&	4.98	&	0.04	&	4.80	&	0.03	&	4219	&	4.68	&	0.65	&	4.73	\\
581	&	19.61	&	0.15	&	1.43	&	0.26	&	1.03	&	0.04	&	0.95	&	0.02	&	8194	&	3.97	&	2.04	&	5.61	\\
587	&	18.92	&	0.54	&	4.76	&	0.07	&	4.27	&	0.07	&	4.13	&	0.07	&	4920	&	4.61	&	0.81	&	5.81	\\
596	&	16.06	&	0.11	&	1.76	&	0.03	&	1.69	&	0.02	&	1.65	&	0.02	&	7831	&	4.17	&	1.72	&	6.85	\\
600	&	19.35	&	0.75	&	7.37	&	0.09	&	6.79	&	0.09	&	6.53	&	0.09	&	3490	&	4.91	&	0.38	&	5.68	\\
604	&	17.99	&	0.2	&	5.04	&	0.04	&	4.54	&	0.03	&	4.39	&	0.03	&	4628	&	4.64	&	0.74	&	3.34	\\
608	&	18.58	&	0.33	&	3.91	&	0.04	&	3.61	&	0.07	&	3.58	&	0.04	&	5648	&	4.53	&	0.98	&	5.92	\\
618	&	23.08	&	0.1	&	2.67	&	0.03	&	2.51	&	0.03	&	2.46	&	0.02	&	6659	&	4.30	&	1.32	&	4.77	\\
632	&	19.06	&	0.07	&	0.73	&	0.32	&	0.78	&	0.21	&	0.65	&	0.02	&	8056	&	3.86	&	2.15	&	5.77	\\
635	&	18.54	&	0.37	&	3.97	&	0.05	&	3.64	&	0.05	&	3.57	&	0.05	&	5622	&	4.53	&	0.98	&	3.24	\\
645	&	15.28	&	1.15	&	5.83	&	0.17	&	5.15	&	0.17	&	4.96	&	0.16	&	4098	&	4.70	&	0.62	&	7.20	\\
663	&	20.77	&	0.76	&	7.47	&	0.08	&	6.84	&	0.08	&	6.62	&	0.08	&	3472	&	4.92	&	0.36	&	5.30	\\
674	&	14.88	&	0.81	&	6.63	&	0.12	&	6.01	&	0.12	&	5.77	&	0.12	&	3703	&	4.80	&	0.49	&	7.39	\\
676	&	17.41	&	0.4	&	3.90	&	0.05	&	3.60	&	0.08	&	3.51	&	0.05	&	5682	&	4.52	&	0.99	&	6.32	\\
679	&	24.62	&	0.74	&	6.36	&	0.07	&	5.74	&	0.07	&	5.48	&	0.07	&	3815	&	4.76	&	0.54	&	4.47	\\
686	&	16.02	&	1.25	&	5.80	&	0.17	&	5.15	&	0.17	&	4.94	&	0.17	&	4111	&	4.70	&	0.62	&	6.87	\\
688	&	14.23	&	0.4	&	3.18	&	0.07	&	2.94	&	0.08	&	2.85	&	0.07	&	6256	&	4.39	&	1.18	&	7.73	\\
697	&	19.43	&	0.12	&	1.50	&	0.02	&	1.44	&	0.03	&	1.38	&	0.03	&	8093	&	4.11	&	1.85	&	5.66	\\
700	&	15.45	&	0.3	&	4.68	&	0.05	&	4.18	&	0.05	&	4.08	&	0.05	&	5000	&	4.61	&	0.82	&	7.12	\\
706	&	15.53	&	1.85	&	5.87	&	0.26	&	5.25	&	0.26	&	5.09	&	0.26	&	4046	&	4.71	&	0.61	&	7.08	\\
718	&	32.43	&	1.5	&	8.53	&	0.10	&	7.96	&	0.10	&	7.68	&	0.10	&	3269	&	5.04	&	0.21	&	3.39	\\
719	&	23.46	&	1.57	&	7.14	&	0.15	&	6.52	&	0.15	&	6.27	&	0.15	&	3550	&	4.88	&	0.42	&	4.69	\\
724	&	25.99	&	3.31	&	6.15	&	0.28	&	5.44	&	0.28	&	5.23	&	0.28	&	3936	&	4.73	&	0.58	&	4.23	\\
\end{longtable}
\end{landscape}
\end{longtab}
}

\end{document}